\documentclass[submission,copyright,creativecommons]{eptcs}
\usepackage{amsfonts}
\usepackage{underscore}
\providecommand{\event}{Infinity 2011} 
\usepackage{breakurl}             
\def\arrow#1{\stackrel{#1}{\longrightarrow}}
\def\inp#1{^{\bullet}{#1}}
\def\out#1{{#1}^{\bullet}\!}
\def\exec#1{{[#1\rangle}}
\def\allprocs{{\cal C}}
\def\supervisors{{\cal S}}
\def\url#1{\sf{#1}}

\def\procs{\Pi}
\def\proc{\pi}
\def\stut{\mathit{stutcl}}
\def\last{\mathit{last}}
\def\vis{\mathit{vis}}
\def\good{\mathit{good}}
\def\en{\mathit{en}}
\def\pref{\mathit{pref}}

\def\procc{\mathit{proc}}
\def\ct{\mathit{ct}}
\def\uc{\mathit{uc}}
\def\df{\mathit{df}}
\def\ngb{\mathit{ngb}}
\def\own{\mathit{own}}

\def\reach{\mathit{reach}}
\def\execc{\mathit{exec}}
\def\attr{\mathit{attr}}
\def\supp{\mathit{supp}}

\title{Practical Distributed Control Synthesis\thanks{The research 
was funded by
Israeli Science Foundation (ISF) grant 1252/09 and by the EPSRC grant EP/H046623/1.}}
\author{Doron Peled
\institute{Department of Computer Science}
\institute{Bar Ilan University}
\institute{Ramat Gan 52900, Israel} \and
       Sven Schewe
\institute{Department of Computer Science}
\institute{University of Liverpool}
\institute{Liverpool, UK}}
\def\titlerunning{Practical Distributed Control Synthesis}
\def\authorrunning{Doron Peled and Sven Schewe}

\usepackage{eepic}
\usepackage{pslatex}
\usepackage{graphicx}
\usepackage{latexsym}
\usepackage{amsthm}
\newtheorem{definition}{Definition}
\newtheorem{lemma}{Lemma}

\begin{document}
\maketitle
\begin {abstract}
Classic distributed control problems have an interesting dichotomy: 
they are either trivial or undecidable.
If we allow the controllers to fully synchronize, then synthesis is trivial. In
this case, controllers can effectively act as a single controller with complete information, resulting in a trivial control problem.
But when we eliminate communication and restrict the supervisors to locally available information, the problem becomes undecidable.
In this paper we argue in favor of a middle way.
Communication is, in most applications, expensive, and should hence be
minimized.
We therefore study a solution that tries to communicate only scarcely and, while allowing communication in order to make joint decision, favors local decisions over joint decisions that require communication.
\end{abstract}

\thispagestyle{plain}

\section{Introduction}

Synthesizing code directly from a formal specification is highly intractable.
Although automated synthesis is an attractive concept, neither is 
the practice of programming currently under threat of extinction, 
nor is automatic synthesis close to become a major factor in code generation. 
Still, some small critical tasks or protocols may be quite tricky for a 
programmer to produce and can greatly benefit from either fully automatic 
synthesis or a computer assisted development methodology.
Prominent representatives of such tasks are concurrency control protocols that 
guarantee mutual exclusion, locking, or efficient memory access. 
The most challenging programming problems are often concurrent in nature, and, alas,
synthesis of concurrent algorithms is undecidable~\cite{PR}.

This undecidability result on synthesizing concurrent
code provides an important information about how {\em not} to attack
the synthesis problem: through a general catch-all algorithmic method.
One common practice to deal with an undecidable result
is to restrict the generality of the problem. This can be done by
limiting the architecture of the system~\cite{PR,KV,MT,FS,SF1,SF2}.
Positive results, however,
are restricted to very limited architectures,
such as pipelines, rings, or assumption about the hierarchy of memory access.

Another approach is to use a heuristic method, accepting that it may
not succeed in all cases. 
A genetic search among the space of
syntactically limited programs, which mutates existing candidates
and progresses based on ranking provided by model checking, is 
described in~\cite{KP1}.
Instead of using a direct synthesis algorithm, this technique generates 
candidate solutions, evaluates their quality (the model checking is 
generalized to a fitness function that estimates the distance from a 
solution), and adjusts them to fitter solutions.
This method is successful in automatically
finding solutions to mutual exclusion~\cite{KP1} and leader election
problems~\cite{KP2} and was even used to detect and
correct an error in a complicated communication protocol~\cite{KP3}.
In principle, such heuristic search techniques can be fully automatic,
though they require human interaction, through setting
the parameters or adjusting them after an unsuccessful run, to be efficient.

We concentrate on synthesizing distributed control~\cite{RR2000,RudiW92,YL}. 
Synthesis is achieved in an incremental way: an 
already existing distributed system is modified to satisfy an 
additional property.
In our case, an invariant. Controlling the system is done by 
selectively blocking transitions. Ideally, local decisions can be
taken by the processes themselves, or equivalently, by supervisors
(one per process) that control the processes and synchronize with them.
It turns out that the controllability problem (whether such distributed
control exists) is also undecidable~\cite{Thistle,Tripakis04}, even for
simple safety properties such as execution according to priorities~\cite{GPQ}.

To challenge this undecidability result, we relax the problem and allow
additional temporary interactions between processes in order
to allow them to acquire sufficient information to decide together on allowing (the converse of blocking) a transition.
Formally, this coordination is mapped to a supervisor.
A variant of this method is to partition the processes into groups of communicating processes, or, likewise, to introduce regional supervisors and assign each process to one of them.
These (regional) supervisors collect enough process information to make control
decisions.
Under this assumption, all processes may, at the limit, interact to
decide globally on the execution of each transition. This reduces the
problem, in the limit, to a sequential control problem, which is trivial for finite state systems.
The efficiency of this method
depends on the amount of additional synchronization needed to enforce
the desired invariant.

The method we use to enforce control is based on {\em knowledge}~\cite{FHMV,M}.
Intuitively, in a distributed system, the knowledge of a process includes
all properties that globally hold in all states consistent with
the local view of the process. It reflects limited visibility of processes
about the situation in other processes. The definition of knowledge is
quite subtle, as it involves some assumptions about the view of a process.
Indeed, in order to make a distributed control decision, a process (or
a supervisor process synchronized with it) must make a choice that
is good for all possible global states that are consistent with
its local view. As process knowledge may not be sufficient,
interaction between processes may be used to acquire the joint knowledge
of several processes. Furthermore, knowledge can be refined based on the
history of an execution. In this way, the number of possible global
states that are consistent with the local view may be reduced, based on
different histories. On the other hand, using this kind of knowledge 
requires the support of an expensive program transformation. We will
discuss at length the use of knowledge in constructing control
for distributed systems.

The knowledge based control synthesis~\cite{M,BBPS,BBGPQ,GPQ}
restricts the executions of the system. 
The information gathered during the model checking stage is used
as a basis for a program transformation that controls the execution of the
system by adding constraints on the enabledness of transitions.
This does not produce new program executions or deadlocks and, consequently, preserves all stuttering closed~\cite{PW} linear 
temporal logic properties of the system~\cite{MP} when no fairness is
assumed.

\section{Preliminaries}

We chose Petri Nets as our model because of the intuitive and concise representation offered by them.
But the method and algorithms developed extend to other models, such as
transition systems, communicating automata, etc.

\begin{definition}
\label{petrinet}
A  {\em (1-safe) Petri Net} $N$ is a tuple
$( P, T, E, s_0 )$ where
\begin{itemize}
\item $P$ is a finite set of {\em places}, 
\item the {\em states} are defined as $S = 2^{P}$ where
$s_0 \in S$ is the {\em initial state},
\item $T$ is a finite set of {\em transitions}, and
\item $E \subseteq ( P \times T ) \cup ( T \times P )$ is a bipartite 
         relation between the places and the transitions.
\end{itemize}
For a transition $t \in T$, we define the set of {\em input places}
$\inp{t}$ as $\{ p \in P \mid ( p, \, t ) \in E \}$,
and {\em output places} $\out{t}$ as $\{ p \in P \mid ( t , \, p ) \in E \}$.
\end{definition}
\begin{definition}
A transition $t$ is {\em enabled} in a state $s$, denoted $s \exec{t}$,
if $\inp{t} \subseteq s$ and $\out{t} \cap s \subseteq \inp{t}$.
A state $s$ is in {\em deadlock} if there is no enabled transition from it.
\end{definition}
\begin{definition}
A transition $t$ can be {\em fired} (or {\em executed})
from state $s$ to state $s'$,
denoted by $s \exec{t} s'$, when 
$t$ is enabled at $s$. Then, $s' = (s \setminus \inp{t}) \cup
\out{t}$. 
\end{definition}

\begin{definition}
\label{independent}
Two transitions $t_1$ and $t_2$ are {\em dependent} if 
$(\inp{t_1} \cup \out{t_1}) \cap (\inp{t_2} \cup \out{t_2}) \neq \emptyset$.
Let $D \subseteq T \times T$ be the {\em dependence} relation.
Two transitions are {\em independent} if they are not dependent.
\end{definition}

\begin{figure}[t]
\begin{center}
\setlength{\unitlength}{0.00045in}
\begingroup\makeatletter\ifx\SetFigFont\undefined%
\gdef\SetFigFont#1#2#3#4#5{%
  \reset@font\fontsize{#1}{#2pt}%
  \fontfamily{#3}\fontseries{#4}\fontshape{#5}%
  \selectfont}%
\fi\endgroup%
{
\begin{picture}(2794,2919)(0,-10)
\put(434,1446){\ellipse{450}{450}}
\put(431,340){\ellipse{450}{450}}
\put(2231,1465){\ellipse{450}{450}}
\put(2231,340){\ellipse{450}{450}}
\put(1349,1407){\ellipse{450}{450}}
\put(450,2577){\ellipse{450}{450}}
\put(2231,2590){\ellipse{450}{450}}
\put(2231,2590){\blacken\ellipse{230}{230}}
\put(2231,2590){\ellipse{230}{230}}
\put(450,2577){\blacken\ellipse{230}{230}}
\put(450,2577){\ellipse{230}{230}}
\put(1349,1407){\blacken\ellipse{230}{230}}
\put(1349,1407){\ellipse{230}{230}}
\path(225,867)(675,867)
\path(2025,1992)(2475,1992)
\path(2025,867)(2475,867)
\path(450,2307)(450,1992)
\path(420.000,2112.000)(450.000,1992.000)(480.000,2112.000)
\path(450,1992)(450,1677)
\path(420.000,1797.000)(450.000,1677.000)(480.000,1797.000)
\path(450,1227)(450,822)
\path(420.000,942.000)(450.000,822.000)(480.000,942.000)
\path(2250,2352)(2250,1947)
\path(2220.000,2067.000)(2250.000,1947.000)(2280.000,2067.000)
\path(2250,1947)(2250,1722)
\path(2220.000,1842.000)(2250.000,1722.000)(2280.000,1842.000)
\path(2250,1227)(2250,867)
\path(2220.000,987.000)(2250.000,867.000)(2280.000,987.000)
\path(450,867)(450,552)
\path(420.000,672.000)(450.000,552.000)(480.000,672.000)
\path(2250,867)(2250,552)
\path(2220.000,672.000)(2250.000,552.000)(2280.000,672.000)
\dottedline{45}(1350,2892)(1350,1632)
\dottedline{45}(1350,1182)(1350,12)
\path(225,1992)(675,1992)
\path(1170,1587)(1169,1589)(1168,1593)
	(1165,1600)(1161,1611)(1155,1625)
	(1147,1645)(1138,1668)(1127,1694)
	(1115,1724)(1101,1756)(1087,1790)
	(1072,1825)(1056,1860)(1039,1896)
	(1022,1931)(1004,1966)(986,2000)
	(967,2033)(947,2065)(926,2096)
	(905,2125)(882,2152)(859,2177)
	(834,2199)(810,2217)(782,2232)
	(756,2240)(731,2242)(709,2239)
	(689,2232)(671,2221)(654,2207)
	(639,2190)(625,2171)(611,2150)
	(599,2128)(588,2106)(577,2084)
	(568,2063)(560,2044)(553,2028)
	(548,2014)(540,1992)
\path(552.815,2115.028)(540.000,1992.000)(609.203,2094.523)
\path(1485,1587)(1486,1589)(1487,1592)
	(1491,1598)(1495,1608)(1502,1621)
	(1511,1639)(1521,1660)(1534,1684)
	(1548,1712)(1564,1743)(1581,1775)
	(1600,1809)(1619,1844)(1638,1880)
	(1659,1916)(1679,1951)(1700,1986)
	(1722,2020)(1744,2053)(1766,2086)
	(1789,2117)(1812,2147)(1836,2175)
	(1861,2201)(1885,2225)(1910,2245)
	(1935,2262)(1962,2276)(1987,2283)
	(2010,2285)(2030,2282)(2047,2275)
	(2063,2264)(2076,2250)(2088,2233)
	(2099,2214)(2109,2193)(2117,2171)
	(2125,2148)(2132,2125)(2138,2102)
	(2143,2079)(2148,2059)(2151,2040)
	(2154,2025)(2157,2012)(2160,1992)
\path(2112.531,2106.222)(2160.000,1992.000)(2171.867,2115.123)
\path(585,867)(585,865)(586,862)
	(588,855)(590,845)(593,832)
	(598,816)(603,798)(609,778)
	(616,756)(624,735)(633,714)
	(643,693)(654,675)(667,658)
	(681,644)(696,633)(714,625)
	(734,621)(757,622)(782,629)
	(810,642)(835,658)(859,678)
	(884,701)(908,726)(931,753)
	(954,781)(976,811)(997,842)
	(1018,873)(1039,906)(1059,939)
	(1078,972)(1097,1005)(1115,1037)
	(1132,1069)(1148,1099)(1163,1127)
	(1176,1152)(1187,1173)(1197,1191)
	(1204,1205)(1215,1227)
\path(1188.167,1106.252)(1215.000,1227.000)(1134.502,1133.085)
\path(2160,822)(2160,821)(2159,818)
	(2157,814)(2154,807)(2150,797)
	(2145,786)(2139,772)(2132,756)
	(2124,740)(2115,723)(2105,705)
	(2094,688)(2083,672)(2070,657)
	(2057,644)(2042,633)(2026,624)
	(2008,617)(1989,614)(1967,614)
	(1943,619)(1918,628)(1890,642)
	(1864,660)(1838,681)(1812,705)
	(1787,731)(1763,758)(1740,787)
	(1718,817)(1697,848)(1676,879)
	(1656,911)(1636,944)(1617,976)
	(1599,1009)(1581,1041)(1565,1072)
	(1549,1102)(1535,1129)(1522,1153)
	(1511,1174)(1503,1192)(1496,1205)(1485,1227)
\path(1565.498,1133.085)(1485.000,1227.000)(1511.833,1106.252)
\put(1640,1362){\makebox(0,0)[lb]{{\SetFigFont{10}{14.4}{\rmdefault}{\mddefault}{\updefault}$p_7$}}}
\put(735,1362){\makebox(0,0)[lb]{{\SetFigFont{10}{14.4}{\rmdefault}{\mddefault}{\updefault}$p_3$}}}
\put(810,2532){\makebox(0,0)[lb]{{\SetFigFont{10}{14.4}{\rmdefault}{\mddefault}{\updefault}$p_1$}}}
\put(2565,2577){\makebox(0,0)[lb]{{\SetFigFont{10}{14.4}{\rmdefault}{\mddefault}{\updefault}$p_2$}}}
\put(2520,1362){\makebox(0,0)[lb]{{\SetFigFont{10}{14.4}{\rmdefault}{\mddefault}{\updefault}$p_4$}}}
\put(2610,282){\makebox(0,0)[lb]{{\SetFigFont{10}{14.4}{\rmdefault}{\mddefault}{\updefault}$p_6$}}}
\put(810,237){\makebox(0,0)[lb]{{\SetFigFont{10}{14.4}{\rmdefault}{\mddefault}{\updefault}$p_5$}}}
\put(45,1947){\makebox(0,0)[lb]{{\SetFigFont{10}{14.4}{\rmdefault}{\mddefault}{\updefault}$a$}}}
\put(0,822){\makebox(0,0)[lb]{{\SetFigFont{10}{14.4}{\rmdefault}{\mddefault}{\updefault}$c$}}}
\put(2565,1947){\makebox(0,0)[lb]{{\SetFigFont{10}{14.4}{\rmdefault}{\mddefault}{\updefault}$b$}}}
\put(2610,777){\makebox(0,0)[lb]{{\SetFigFont{10}{14.4}{\rmdefault}{\mddefault}{\updefault}$d$}}}
\end{picture}
}

\end{center}
\vspace*{-1.3mm}
\caption{\label{first} A Petri Net}
\end{figure}
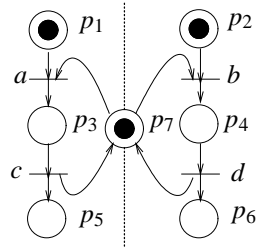

\begin{figure}
\setlength{\unitlength}{0.00045in}
\begin{center}
\begin{picture}(2794,2829)(0,-10)
\put(434,1401){\ellipse{450}{450}}
\put(431,295){\ellipse{450}{450}}
\put(2231,1420){\ellipse{450}{450}}
\put(2231,295){\ellipse{450}{450}}
\put(450,2532){\ellipse{450}{450}}
\put(2231,2545){\ellipse{450}{450}}
\put(2231,2545){\blacken\ellipse{230}{230}}
\put(2231,2545){\ellipse{230}{230}}
\put(450,2532){\blacken\ellipse{230}{230}}
\put(450,2532){\ellipse{230}{230}}
\path(225,1947)(675,1947)
\path(225,822)(675,822)
\path(2025,1947)(2475,1947)
\path(2025,822)(2475,822)
\path(450,2262)(450,1947)
\path(420.000,2067.000)(450.000,1947.000)(480.000,2067.000)
\path(450,1947)(450,1632)
\path(420.000,1752.000)(450.000,1632.000)(480.000,1752.000)
\path(450,1182)(450,777)
\path(420.000,897.000)(450.000,777.000)(480.000,897.000)
\path(2250,2307)(2250,1902)
\path(2220.000,2022.000)(2250.000,1902.000)(2280.000,2022.000)
\path(2250,1902)(2250,1677)
\path(2220.000,1797.000)(2250.000,1677.000)(2280.000,1797.000)
\path(2250,1182)(2250,822)
\path(2220.000,942.000)(2250.000,822.000)(2280.000,942.000)
\path(450,822)(450,507)
\path(420.000,627.000)(450.000,507.000)(480.000,627.000)
\path(2250,822)(2250,507)
\path(2220.000,627.000)(2250.000,507.000)(2280.000,627.000)
\dottedline{45}(1350,2802)(1350,12)
\put(765,1317){\makebox(0,0)[lb]{{\SetFigFont{10}{14.4}{\rmdefault}{\mddefault}{\updefault}$p_3$}}}
\put(810,2487){\makebox(0,0)[lb]{{\SetFigFont{10}{14.4}{\rmdefault}{\mddefault}{\updefault}$p_1$}}}
\put(2565,2532){\makebox(0,0)[lb]{{\SetFigFont{10}{14.4}{\rmdefault}{\mddefault}{\updefault}$p_2$}}}
\put(2520,1317){\makebox(0,0)[lb]{{\SetFigFont{10}{14.4}{\rmdefault}{\mddefault}{\updefault}$p_4$}}}
\put(2610,237){\makebox(0,0)[lb]{{\SetFigFont{10}{14.4}{\rmdefault}{\mddefault}{\updefault}$p_6$}}}
\put(810,192){\makebox(0,0)[lb]{{\SetFigFont{10}{14.4}{\rmdefault}{\mddefault}{\updefault}$p_5$}}}
\put(45,1902){\makebox(0,0)[lb]{{\SetFigFont{10}{14.4}{\rmdefault}{\mddefault}{\updefault}$a$}}}
\put(0,777){\makebox(0,0)[lb]{{\SetFigFont{10}{14.4}{\rmdefault}{\mddefault}{\updefault}$c$}}}
\put(2565,1902){\makebox(0,0)[lb]{{\SetFigFont{10}{14.4}{\rmdefault}{\mddefault}{\updefault}$b$}}}
\put(2610,732){\makebox(0,0)[lb]{{\SetFigFont{10}{14.4}{\rmdefault}{\mddefault}{\updefault}$d$}}}
\end{picture}
\caption{\label{second} A Petri Nets with priorities $a \ll d$ and $b \ll c$}
\end{center}
\end{figure}
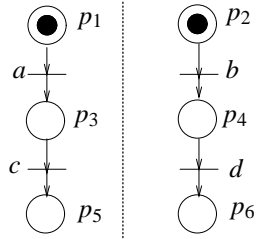

Transitions are visualized as lines,
places as circles, and the relation $E$ is represented using
arrows. In Figure~\ref{first}, there are places
$p_1, \, p_2, \, \ldots , \, p_7$ and transitions
$a, \, b, \, c, \, d$.
We depict a state by putting full circles, called {\em tokens}, 
inside the places of that state. In the
example in Figure~\ref{first}, the initial state $s_0$ is
$\{ p_1 , \, p_2 , \, p_7 \}$.
The transitions
that are enabled from the initial state are $a$ and $b$. 
If we fire transition $a$ from the initial state,
the tokens from $p_1$ and $p_7$ will be removed, and a token
will be placed in $p_3$. 
In this Petri Net, all transitions
are dependent on each other, since they all involve the
place $p_7$. Removing $p_7$, as in Figure~\ref{second},
makes both $a$ and $c$ become independent from both $b$ and $d$.

\begin{definition}
\label{execution}
An {\em execution} of a Petri Net $N$ is a maximal 
(i.e., it cannot be extended) alternating
sequence of states and transitions $s_0 t_1 s_1 t_2 s_2 \ldots$, where
$s_0$ is the initial state,
such that, for each states $s_i$ in the sequence, $s_i \exec{t_{i+1}} s_{i+1}$. 
We denote these executions by $\execc ( N )$.
\end{definition}
For convenience, we sometimes use as executions just the 
sequence of states, or just the sequence of transitions,
as will be clear from the context.
A state is {\em reachable} in a Petri Net if it appears on
at least one of its executions. We denote the reachable states
of a Petri Net $N$ by $\reach(N)$.

We use places also as state predicates.
As usual, we write $s \models p_i$ iff $p_i \in s$ and extend this in 
the standard way to Boolean combinations on state predicates.
For a state $s$, we denote by $\varphi_{s}$ the formula that is a conjunction of the places in $s$ and the negated places not in $s$.
Thus, $\varphi_s$ is satisfied exactly by the state $s$.
For the Petri Net in Figure~\ref{first}, the initial state $s_0$ satisfies 
$\varphi_{s_0} = p_1 \wedge p_2 \wedge \neg p_3 \wedge
\neg p_4 \wedge \neg p_5 \wedge \neg p_6 \wedge p_7$.
For a set of states $Q \subseteq S$, let 
$\varphi_{Q} = \bigvee_{s\in Q} \varphi_s$, or 
any logically equivalent propositional formula, 
be a {\em characterizing formula} of $Q$.
As usual in logic, when $\varphi_Q$ and $\varphi_{Q'}$ characterize sets of
states $Q$ and $Q'$, respectively, then 
$Q \subseteq Q'$ exactly when $\varphi_Q \rightarrow \varphi_{Q'}$.

An invariant~\cite{C} of $N$ is a subset of
the states $Q \subseteq 2^{S}$;  a net $N$ satisfies
the invariant $Q$ if $\reach ( N ) \subseteq Q$.
A {\em generalized invariant} of $N$ is a set of pairs
$I \subseteq S \times T$; a net $N$ satisfies $I$ if,
whenever $s \exec{t}$ for a reachable~$s$, then $( s, t ) \in I$.
This covers the above simple case of an invariant
by pairing up every state that appears in $Q$ with {\em all} transitions~$T$.

\begin{definition}
\label{restricted}
An execution of a Petri Net $N$ {\em restricted} with respect to 
a set $I \subseteq S \times T$, denoted $\execc_I (N)$, is a maximal 
set of executions $s_0 t_1 s_1 t_2 s_2 \ldots\in \execc ( N )$ such that,
$s_0$ is the initial state,
for each states $s_i$ in the sequence, $s_i \exec{t_{i+1}} s_{i+1}$, 
and furthermore $(s_i,t_{i+1})\in I$.
The set of states reachable in $\execc_I (N)$ is denoted $\reach_I ( N )$.
\end{definition}

\begin{definition}
For a set of executions $X$,
let $\mathit{pref} ( X )$ be the set of prefixes (including
full executions) of $X$.
\end{definition}
Denote the last state of a finite prefix $h$ of an execution by
$\last ( h )$.

\begin{lemma}
$\reach_I ( N ) \subseteq \reach ( N )$ and
$ \mathit{exec}_I ( N  ) \subseteq \pref (\execc ( N ))$.
\end{lemma}

As an example of a property we may want to enforce, consider
prioritized executions.
\begin{definition}
A {\em Petri Net with priorities} is a pair $( N, \ll )$ with
$N$ a Petri Net and $\ll$ a partial order relation among
the transitions $T$ of $N$.
\end{definition}
Let $I_\ll = \{(s,t)\mid 
\mbox{$s \exec{t}$ and $\forall t' \in T \, s\exec{t'} \rightarrow t' \ll
t$} \}$.
The set of \emph{prioritized executions} $\execc_{I_\ll}(N)$ of $( N, \ll )$ is the set of executions restricted to $I_\ll$.
The executions of the Petri Net $M$ in Figure~\ref{second} (when
the priorities $a \ll d$ and $b \ll c$
are {\em not} taken into account) include
$a b c d, a c b d , b a c d, b a d c$,
etc. However, the prioritized executions of $(M,\ll)$ are the same as the
executions of the Net $N$ in Figure~\ref{first}.

\begin{definition}
\label{procs}
A {\em process} $\proc$ of a Petri Net $N$ is a subset of
the transitions $T$.
\end{definition}

We will represent the separation of
transitions of a Petri Net into
processes using dotted lines.
We assume a given set of processes $\allprocs$ that {\em covers} all transitions
of the net, i.e., $\bigcup_{\proc \in \allprocs} \proc = T$.
A transition can belong to several processes, e.g., when it
models a synchronization between processes. 
Let $\procc ( t ) = \{ \pi \mid t \in \pi \}$
be the set of processes to which $t$ belongs.
For the Petri Net in Figure~\ref{first}, there are two executions:
$a c b d$ and $b d a c$. 
There are two processes: the {\em left} process $\proc_l = \{ a, c \}$ 
and the {\em right} process $\proc_r = \{ b , d \}$.

The {\em neighborhood} of a set of processes $\procs$ includes all 
places that are either
inputs or outputs to transitions of $\procs$.
\begin{definition}
The {\em neighborhood} $\ngb ( \proc )$ of a process $\proc$ is the 
set of places $\bigcup_{t \in \proc} (\inp{t} \cup \out{t})$.
For a set of processes $\procs \subseteq \allprocs$, $\ngb ( \procs ) = 
\bigcup_{\proc \in \procs} \ngb ( \proc )$.
\end{definition}

A set of processes $\procs$ {\em owns} the places in their neighborhood
that can gain or lose a token by a transition $t$ only if $t$ is
{\em exclusively} in $\procs$.
\begin{definition}
The set of places {\em owned} by a set of processes (including a singleton
process) $\procs$, denoted $\own ( \procs )$, is 
$\ngb ( \procs ) \setminus  \ngb ( \allprocs \setminus \procs )$.
\end{definition}


When a notation refers to a set of processes $\procs$,
we will often replace writing the singleton
process set $\{ \proc \}$ by writing $\proc$,
e.g., we write $\own ( \proc )$. Note that 
$\ngb (\procs_1 ) \cup \ngb (\procs_2 ) = \ngb ( \procs_1
\cup \procs_2 )$, while 
$\own (\procs_1 ) \cup \own (\procs_2 ) \subseteq \own ( \procs_1
\cup \procs_2 )$. 
The neighborhood of process $\proc_l$ in the Petri Net of
Figure~\ref{first} is
$\{ p_1 , p_3 , p_5 , p_7 \}$.
Place $p_7$  is neither owned by $\proc_l$, nor by $\proc_r$, but it is 
owned by $\{ \proc_l , \proc_r \}$.
It belongs to the neighborhood of both processes and acts
as a semaphore. It can be captured by the execution of $a$ or
of $b$, guaranteeing that
$\neg ( p_3 \wedge p_4 )$ is an invariant of the system.

Our goal is to control the system to satisfy a generalized
invariant by restricting some of
its transitions from some of the states. 
The setting of the control problem may impose that 
only part of the transitions, $\ct ( T ) \subseteq T$, 
called {\em controllable} transitions, can be selectively supported 
by the processors that contain it.
(It blocks if no processor supports it.) 
The other transitions,
$\uc ( T ) = T \setminus \ct ( T )$, are {\em uncontrollable}.
Note that we may be at some state where either some uncontrollable transitions, 
or all enabled transitions,
violate the generalized invariant. Being
in such states is therefore ``too late''; part of the
controlling task is to avoid reaching such states.

In control theory, the transformation that takes a system and
allows blocking some transitions adds a
supervisor process~\cite{RW}, which is usually an automaton
that runs {\em synchronously} with the controlled system. 
This (finite state) automaton observes the controlled system, progresses
according to the transitions it observes, and
blocks some of the enabled transitions, depending on its current
state. In a similar way, in
distributed control~\cite{YL,RudiW92,RR2000}, for each process
we assign such a supervisor, which changes its states each time
the process it supervises makes a transition, or when a visible transition
of another process (e.g., through the change of shared variables)
is executed. Based on its states, the supervisor allows (supports)
transitions of the controlled process. In a disjunctive control
architecture~\cite{YL},
if no supervisor suports an,
otherwise enabled, transition, it cannot execute and is thus blocked. Such
a supervisor can be amalgamated, through a transformation, into the
code of the controlled process. In order to capture this for
Petri Nets, without a complicated transition splitting
transformation, we use an extended model, as defined below. In particular,
it allows adding enabling conditions and variable transformation to capture
the encoding of the local supervision of the processes. It would also
allow encoding additional asynchronous supervision in our solution.

\begin{definition}
\label{extended}
An {\em extended Petri Net}~\cite{K} is a 
Petri Net with a finite set of variables $V_{\proc}$ over a finite domain per
each process $\proc \in \procs$.
In addition, a transition $t$ can be augmented with a predicate 
$\en_t$ on the variables
$V_t = \cup_{\proc \in \procc ( t )} V_\proc$
and a transformation function $f_t ( V_t )$. In order
for $t$ to fire, $\en_t$ must hold in addition to the basic 
Petri Net enabling condition on the input and output places of $t$. 
When $t$ fires, in addition to the usual changes to the tokens,
the variables $V_t$ are updated according to the transformation~$f_t$.
\end{definition} 

We transform a Petri Net $N$ and a generalized invariant 
$I$ into an extended
Petri Net $N'$ that allows only the executions of $N$ controlled 
to satisfy $I$. 
\begin{definition}
\label{transformation}
A {\em controlling} transformation obeys the 
following conditions:
\begin{itemize}
\item New transitions and places can be added. 
\item The input and output places
of the new transitions are disjoint from the existing places.
\item Variables, conditions and transformations
can be added to existing transitions.
\item
Existing transitions will remain with the same input and output places.
\item 
It is not possible to fire from some point an 
infinite sequence consisting of only added transitions.
\end{itemize}
\end{definition}
Added transitions
are grouped into new (supervisory) processes.
Added  variables will represent some knowledge-dependent 
finite memory for controlling the system, and some
interprocess communication media between the original processes and
the added ones.
Processes from the original net will have disjoint sets of
variables from one another. The independence between the 
original transitions
is preserved by the transformation, although some coordination may
be enforced indirectly through the interaction with 
the new supervisory processes.

\begin{definition} 
Let $s \lceil_{\allprocs}$ map a state $s$ of the 
transformed version $N'$ into
the places of the original version $N$ 
by projecting out 
additional variables and places that $N'$ may have on 
top of the places of $N$. This definition is also extended to executions (as
sequences of states).
\end{definition}
This projection allows us to relate the sets of
states of the original and transformed version. 
Firing of a transitions added by the controlling transformation
does not change
$s \lceil_{\allprocs}$ and is not considered to violate $I$
(the requirement that $( s_i, t_{i+1})$ in Definition~\ref{restricted}
is imposed only when $t_{i+1}$ is from the original net $N$).
Note that our restrictions on the
transformation implies that the sets $\ngb ( \procs )$ and $\own ( \procs )$
for $\procs \subseteq \allprocs$ are not affected by the transformation.
Furthermore, albeit the rich structure of extended Petri Nets, our
control transformation will allow a finite state control for a finite state
system.

\begin{definition}
Two executions $\sigma$ and $\sigma'$, viewed as sequences
of states, are equivalent up to stuttering~\cite{PW} when, by replacing
any finite adjacent repetition of the same state by a single occurrence
in both $\sigma$ or $\sigma'$,
we obtain the same sequence. Let $\stut ( \Gamma )$ be the
stuttering closure of a set $\Gamma$ of sequences, i.e., all sequences
that are stuttering equivalent to some sequences in $\Gamma$.
\end{definition}

\begin{lemma}
\label{XXX}
A controlling transformation produces an extended Petri Net
$N'$ from $N$ such that $\execc ( N' ) \lceil_{\allprocs} \subseteq 
\pref (\stut ( \execc ( N ) ) ) $.
\end{lemma}

The controlling transformation may introduce new deadlocks,
hence the lemma above asserts about the prefixes of the
original executions. Of course, this is not a desirable outcome
of the control transformation, and the solutions that will
be given to the distributed control problem will circumvent it.

\section{Process Knowledge and Joint Process Knowledge}

The knowledge of a process 
at a given execution point
consists of facts that hold in all global
states that are consistent with the current local view of this process.
The current local view represents the limited ability of a process
to observe the global state of the system. A process may be aware 
of its own local variables and shared variables in its neighborhood.
Similarly, we can define the
joint knowledge of several processes, by considering their
joint local view.

According to the limited observability of
the processes $\procs$, 
we can define an equivalence
relation $\equiv_{\procs} \subseteq S \times S$ 
(when the set of processes
$\procs$ is a singleton, we can write $\equiv_{\proc}$) 
among the states $S$ of the system;
if the current state is $s \in S$, then the processes
$\procs$ cannot distinguish, given their joint local view,
between $s$ and any state equivalent to it according to
$\equiv_{\procs}$. Such an equivalence relation is the
basis of the definition of knowledge~\cite{FHMV}.

\begin{definition}
\label{knowledge}
The processes ${\procs}$ (jointly) {\em know} a property $\psi$ 
in a state $s$, denoted $s \models K_{\procs} \psi$, if,
for all $s'$ such that  $s \equiv_{\procs}  s'$,
we have that $s' \models \psi$.
\end{definition}

In the Petri Nets model,
the equivalence relation $\equiv_{\procs}$ can be
defined by restricting first each state to a part of a state. Then,
states that share the same part are considered equivalent.
There are several possibilities to restrict the part of 
a state that is associated with a subset of the processes
$\procs$. We will give two possibilities for such a restriction.
The first one is that of {\em local information}, which takes
the part of the state that includes the neighborhood of
the processes $\procs$. This Petri Nets definition
corresponds, in general systems, to the variables that
can be read or written by the processes $\procs$.
The second such restriction is
that of {\em local state} (different names were chosen only
to make a distinction), based on restricting states to the
places that the processes $\procs$ own. This corresponds, in
general systems, to the variables that only the processes
$\procs$, and no other processes, can change (write).

\begin{definition}
\label{local}
The {\em local information} of a set of processes $\procs$ of a Petri Net $N$
in a state $s$ is $s\lceil_{\procs} = s \cap \mathit{nbg} ( \procs )$.
\end{definition}
In the Petri Net in Figure~\ref{first}, the local information of
$\proc_l$ in any state $s$ consists of the restriction of $s$
to the places $\{ p_1 , p_3 , p_5, p_7 \}$. 
In the depicted initial state, the local information is $\{ p_1 , p_7 \}$.
\begin{definition}
\label{localstate}
The {\em local state} of a set of processes $\procs$ of a Petri Net $N$
in a state $s$ is $s\lfloor_{\procs} = s \cap \own ( \procs )$.
\end{definition}
It is always the case that $s\lfloor_{\procs} \subseteq s\lceil_{\procs}$.
The local state of $\proc_l$
in the initial state of Figure~\ref{first} is $\{ p_1 \}$. 

\begin{lemma}
\label{update}
If $\proc \not\in \procs$ then
$s \lfloor_{\procs \cup \{ \proc \}}$ is the (disjoint) union of
$s \lfloor_{\procs}$ and
$s \lceil_{\proc} \cap \own ( \procs \cup \{ \proc \} )$.
\end{lemma}

In the following definitions, we can often use either the local 
information or
the local state. When this is the case, we will use $s |_{\procs}$
instead of either $s \lceil_{\procs}$ or $s \lfloor_{\procs}$. 

\begin{definition}
\label{haha}
Let $\procs \subseteq \allprocs$ be a set of processes.
Define an equivalence relation 
$\equiv_{\procs} \subseteq \reach(N) \times \reach(N)$ such that
$s \equiv_{\procs} s'$ when $s|_{\procs} = s' |_{\procs}$.
\end{definition}
As $s |_\procs$ can stand for either $s \lceil_{\procs}$ or
$s \lfloor_{\procs}$, this gives two different equivalence relations.
When it is important to distinguish between them, we denote the
one based on ``$\lceil$'' as $\equiv^{w}_{\procs}$ (weak equivalence) and the one
based on ``$\lfloor$'' as $\equiv^{s}_{\procs}$ 
(strong equivalence).
\begin{lemma}
\label{execo}
If $t \in \proc$ and $s \equiv^{w}_{\proc} s'$ then
$s \exec{ t }$ if, and only if, $s' \exec{ t }$.
\end{lemma}
That is, the enabledness of a transition depends only on the local 
information of a process that contains it.
This does not hold when we replace $\equiv^{w}_{\proc}$
by $\equiv^{s}_{\proc}$.
In the Prioritized Petri Net 
in Figure~\ref{first}, e.g., we have
that  
$\{ p_1, p_2 , p_7\} \equiv^{w}_{\proc_l} \{ p_1 , p_4 , p_7\}$, 
since $\proc_l$ has the same local information $\{ p_1 , p_7\}$ in both states.
The state $\{ p_1, p_4 \}$ is not equivalent to either of these
states. On the other hand, these three states are equivalent
according to $\equiv^{s}_{\proc_l}$ ($p_7$ is not in $own ( \proc_l )$).

Corresponding with the two equivalence relations
of Definition~\ref{haha}, we distinguish between knowledge based 
on strong equivalence $\equiv_{\procs}^s$ (and hence on local states), 
denoted $K^{s}_{\procs} \varphi$ and
knowledge based on weak equivalence $\equiv_{\procs}^w$
(and hence local information), denoted $K^{w}_{\procs} \varphi$. 
The knowledge based on the local state 
(resp.\ local information) is called 
{\em s}trong (resp.\ {\em w}eak) knowledge.
Since the local information determines the local state (while multiple local states may have the same local information), we have
$K^{s}_{\procs} \varphi \rightarrow K^{w}_{\procs} \varphi$.
Consequently, we may {\em know} more under weak knowledge.

The motivation for the different definitions of equivalence and,
subsequently, the different definitions of knowledge is as follows.
In order to make choices (to support or block a transition)
that take into account knowledge based on local
information, a process, or a set of processes, needs to have some guarantee
that the local information will not be changed by other processes while
it is collecting information from the processes or
making the decision. For a single process, this may be achieved 
by the underlying hardware. But it is 
unreasonable to require such a guarantee for a set of processes that
either temporary interact (interactions take time and
other processes
may meanwhile progress) or send their current local view to some supervisor
process that collects views from several processes.
Thus, for decisions involving a set of processes, 
strong knowledge, based on the joint local state, is used instead.

The classical definition of knowledge is based 
on relations $\equiv_{\procs}$
over the reachable states $\mathit{reach}_I ( N )$. 
However, when using knowledge to control a
system to satisfy a generalized invariant, one may calculate the
equivalences and the knowledge based on the states 
$\mathit{reach}_I ( N )$ that appear in the executions
of the original system that satisfy this generalized invariant $I$.
This (cyclic looking) claim is proved~\cite{BBGPQ} by induction on the progress
of the execution in the controlled system: for a state
already on such an execution (by the inductive assumption) the controlled 
system allows firing only transitions that preserve the
generalized invariant, hence is also
in $\mathit{reach}_I ( N )$. We may
need to restrict the generalized invariant $I$, in order
not to introduce new deadlocks. This means even fewer reachable states,
which can consequently increase the knowledge further.

One of the main challenges of using knowledge for controlling systems
is that it is not always possible to decide, based on the local
(or joint) knowledge, whether or not allowing a transition will guarantee
the desired generalized invariant. One tool that can be used in this
case is to allow additional interactions between processes, or knowledge
accumulation by additional asynchronous supervisors. This will be
\pagebreak[3]
explained later. However, before progressing
to such an expensive solution, we may
also try to improve the knowledge by refining the equivalence relation
that is used in its definition.

The definitions of knowledge that we used assumes that the processes do not
maintain a log with their history. The use of
knowledge with such a log, 
called {\em knowledge with perfect recall}~\cite{M}, is 
discussed in~\cite{BBPS}. 
Consider an equivalence $\approx_{\proc}$ 
between histories that seem undistinguishable to the
process $\proc$.
Two finite prefixes $h$, $h'$ of Petri Net executions will
be considered equivalent for $\approx_{\proc}$ if
the projection of $h$ on transitions visible to $\proc$ are the same in both
$h$ and $h'$. Specifically for Petri Nets, we can define the transitions 
$\vis ( \proc ) = \{ t | ( \inp{t} \cup \out{t}) 
\cap \ngb (\proc) \neq \emptyset \}$ ($t$ is dependent 
on some transitions in $\proc$).
In this case, the last states $\last ( h )$ and
$\last ( h' )$ of $h$ and $h'$, respectively,
are equivalent under $\equiv^w$ (and hence also 
under $\equiv^{s})$. This can be shown by induction over the length
of the prefixes, based on the fact that only the transitions in
$\vis ( \proc )$ affect $\ngb ( \proc ) \supseteq \own ( \proc)$.

\begin{definition}
Let $h \models \psi$ exactly when 
$\last ( h  )\models \psi$.
Then we define \emph{past knowledge}, where
$h \models K^p_{\proc} \psi$ if, for all
$h' \approx_\proc h$, $h \models \psi$. 
\end{definition}

In particular for properties $\psi$ that depend only on the last state of $h$,
the use of the history refines the weak equivalence between states: $h \approx_\proc h'$ implies $\last ( h ) \equiv^w_\procs \last ( h' )$.
To take advantage of the refined definition of knowledge,
we need somehow to distinguish local states that have non equivalent
histories. On the face of it, this seems to require unbounded memory.
However, looking deeper into the new definition of knowledge, one
can observe that the following finite construction will work~\cite{M,BBPS}.

\begin{definition}
Let $\mathit{\bigtriangleup}_{\proc}$ be
the set of finite sequences of transitions that do not change
the neighborhood of $\proc$ (i.e., independent with the transitions
in $\proc$). 
\end{definition}

\begin{definition}
Let ${\cal A} = ( S, s_0, T )$ be a finite automaton representing
the global states $S$ of a Petri Net
$N$, including the initial state $s_0 \in S$ and the transitions $T$ 
between them.  
For each process $\proc$, we construct an automaton ${\cal A}_{\proc}$
representing the set of states of ${\cal A}$ where the Petri Net $N$
can be after a given local history.  The automaton ${\cal A}_{\proc}$ has the
following components:
\begin{itemize}
\item
The set of states is $2^{S}$. 
\item
The initial state is the set of states 
$\{ s | \exists \mu \in \mathit{\bigtriangleup}_{\proc} 
\, s.t.\, s_{0} \exec{\mu} s \}$. 
That is, the initial state of this automaton contains all states
obtained from $s_0$ by executing a finite number of
transitions independent of (i.e., invisible to) $\proc$. 
\item
The transition relation is
$\Gamma \arrow{t} \Gamma'$ between two states $\Gamma, \, \Gamma' \in 2^{S}$
and a transition $t \in T$ is as follows:
$\Gamma' = \{ s' | \exists s \in \Gamma \, \exists 
\mu \in \mathit{\bigtriangleup}_{\proc} \,s.t., s \exec{t \mu} s' \}$.
That is, a move from $\Gamma$ to $\Gamma'$  corresponds to
the execution of a transition $t$ that changes
the neighborhood of $\proc$ followed by transitions independent of $\proc$.
\end{itemize}
\end{definition}

Then, one may use $K^p_{\proc} \psi$ instead of $K^w_{\proc}$ for locally
supporting transitions. (Note that $K^w_{\proc} \rightarrow K^p_{\proc}$.)
However, the size of each
such automaton (one per process $\proc$) can be exponential in the
size of the global state space.
Knowledge of perfect recall can be implemented by adding a synchronized
supervisor with memory (basically implementing the automaton
${\cal A}_{\proc}$).
It is natural to ask whether one can make an even finer distinction
between states than with knowledge of perfect recall. This is indeed
possible, but at the cost of a more involved program transformation.
We may augment in our transformation the context of
the interprocess communication between processes with additional
transformation, that would implement the support for additional knowledge.
Such a transformation can, e.g., be based on Gossip Automata~\cite{MS},
providing the most recent past local view of any other process.

We henceforth use knowledge formulas combined with
Boolean operators and propositions. For a detailed
syntactic and semantic description of logics with knowledge
one can refer, e.g., to~\cite{FHMV}.
Once $s \models K_{\procs} \psi$ is defined, $\psi$
can also be a knowledge property, hence
$s \models K_{\procs'} K_{\procs} \psi$ (knowledge about knowledge) is also
defined, though the finite-state representation described above only applies to past knowledge used in outermost knowledge operators.

\begin{lemma}
\label{simpol}
If $s \models K_{\procs} \varphi$ and $s \equiv_{\procs} s'$, then
$s' \models K_{\procs} \varphi$.
\end{lemma}
\begin{lemma}
Knowledge is monotonic with respect to the set of observing processes:
if $\,\procs' \subseteq \procs$ then $K_{\procs'} \varphi \rightarrow 
K_{\procs} \varphi$.
\end{lemma}

\begin{lemma}
\label{theo}
Given that $s \models K_{\procs} \varphi$ in some basic Petri Net $N$,
then $s \models K_{\procs} \varphi$ also in a transformed version~$N'$.
\end{lemma}


Enforcing prioritized executions in a
completely distributed way may be impossible. 
In Figure~\ref{second}, 
$a$ and $c$ belong to the left process $\proc_l$, and
$b$ and $d$ belong to the right process $\proc_r$, with no
interaction between the processes. The left process
$\proc_l$, upon having a token in $p_1$, cannot
locally decide whether to execute $a$; the priorities
dictate that $a$ can be executed if $d$ is not enabled,
since $a$ has a lower priority than $d$. But 
cannot distinguish between the cases where $\proc_r$ has
a token in $p_2$, $p_4$, or $p_6$.

In the Prioritized Petri Net 
in Figure~\ref{second}, e.g., we have
that $\{ p_1, p_2 \} \equiv^{w}_{\proc_l} \{ p_1 , p_4 \}$, 
since in both states $\proc_l$ has the same local information $\{ p_1 \}$.
In the state $\{ p_1 , p_2 \}$, $a$ is a maximal priority enabled
transition (incomparable with $b$), while in $\{ p_1 , p_4 \}$, $a$ is not
maximal anymore, as we have that $a \ll d$, and both $a$ and $d$ are now
enabled. 
In the initial state
the local information (and also the local state) of $\proc_l$ is
$\{ p_1 \}$. Thus, $\proc_l$ does not have enough 
knowledge to support any transition
since $\{ p_1, p_2 \} \equiv^{w}_{\proc_l} \{ p_2 , p_3 \}$). 
Similarly, the local information of $\proc_r$ is
$\{ p_2 \}$, which also is not sufficient to support any transition. 
After they both hang on a supervisor, it has enough information 
to support~$a$~or~$b$.

\section{A Globally Controlled System}
\label{globalgame}

Before providing a solution to the distributed control problem
we need to provide a solution to the related global control
problem. Some reachable states
are not allowed according to the generalized invariant. In order
not to reach these states, resulting in an immediately deadlock,
we may need to avoid some transitions
that lead to such states from previous states. 
This is done using game theoretical search.

The game is played between a \emph{constructor}, who wants to preserve the generalized invariant $I$ indefinitely
(or reach a state that is already a deadlock in the
original system $N$), and a {\em spoiler}, who has the opposite goal.
The game is played on the states $S$ of a net.
It starts from the initial state $s_0$ and ends if a deadlock state is reached (and may go on forever).
In each round, the constructor player chooses a nonempty 
subset of enabled transitions 
that must include all enabled uncontrollable transitions.
Subsequently, the spoiler chooses a transition from this set, which is 
then executed.
The spoiler wins as soon as she can choose a transition that violates $I$,
i.e., $(s,t)\notin I$, while the constructor wins if this condition never holds (on an infinite run or a finite run that ends in a deadlock).

We can define an ``attractor'' $\attr ( A )$ that contains all states in $A$ and all states that the spoiler can force to $A$ in a single transition.
A state $s$ is in $\attr ( A )$ if one of the following conditions holds:
\vspace{-1ex}\begin{itemize}
\item
$s \in A$,
\item
there exists an uncontrollable transition $t \in \uc ( T )$ enabled in
$s$ with $s \exec{t} s'$ and either $s' \in A$, or $( s, t ) \not\in I$, or
\item
$s$ is not a deadlock state in the Petri Net $N$ and, for all transitions $t$ enabled in $s$,
such that $s \exec{t} s'$ and 
$( s , t ) \in I$, it holds that $s' \in A$.
\end{itemize}

As usual, we define $\attr^{n+1} ( A ) = \attr ( \attr^{n} ( A ) )$, where
$\attr^{0} ( A ) = A$. Because of the monotonicity
of the $\attr ( A )$ operator (with respect to set inclusion)
and the finiteness of the state space, there is a least fixpoint
$\attr^{\ast} ( A )$, which is $\attr^{n} ( A ) = \attr^{n+1} ( A )$
for some (smallest) $n$.

Now, let $I_G=\{(s,t) \in I \mid s \exec{t} s' \mbox{ and } s' \notin
\attr^{\ast} ( \emptyset )\}$.
Let $G = \reach_{I_G} ( N ) $ if $s_0 \notin \attr^{\ast} ( \emptyset )$,
otherwise $G = \emptyset$.
These are the ``good'' reachable states in the sense that they are 
allowed by $I$ \emph{and}
the system can be controlled to henceforth adhere to $I$.
\begin{definition}
Let $R = \{ ( s, t ) \in I \mid \exists s'\, s\exec{t}s' \wedge s , s' \in G \}$
be the {\em safe transition relation}.
\end{definition}

If the initial state is good ($s_0 \in G$), then the constructor can 
win by playing according to $R$.
If, on the other hand, $s_0$ is in the 
attractor $\attr^{*} ( \emptyset )$ of the bad states, then $s_0$ is
in $\attr^{n} ( \emptyset )$ for some $n\leq|S|$.
By the definition of $\attr^{n} ( \emptyset )$, the spoiler can 
force the game to  $\attr^{n-1} ( \emptyset )$ in the next step, 
then to  $\attr^{n-2} ( \emptyset )$, and so forth, and thus make
sure the bad states are reached within at most $n$ steps. 

\begin{lemma}
The constructor can force a win if, and only if, $s_0 \in G$.
\end{lemma}

This game can obviously be evaluated quickly on the \emph{explicit} game graph, and hence in time exponentially in the number of places.
EXPTIME completeness can be demonstrated by a simple reduction from 
the {\sf PEEK-$G_5$}~\cite{SC} game~\cite{KPS}.
Deciding if the constructor can force a win is PSPACE complete for Petri
Nets with only controllable transitions~\cite{KPS}.

\subsection*{Model Checking}

We will use the following propositional formulas, with propositions that 
are the places of the Petri Net:
\begin{description}
\item{- The good states $G$:} $\varphi_{G}$.
\item{- The states where a transition $t$ is enabled:} 
$\varphi_{\mathit{en}(t)}$.
\item{- At least one transition is enabled, i.e., 
there is no deadlock:} $\varphi_{\mathit{\df}} =
\bigvee_{t \in T} \varphi_{\mathit{en}(t)}$.
\item \mbox{- Transition $t$ is allowed from the current state by the safe transition relation $R$: $\varphi_{\good (t )}$}
\item{- The local information (resp.\ local state) of processes $\procs$ at state $s$:}
$\varphi_{s \lceil_{\procs}}$ 
(resp.\ $\varphi_{s \lfloor_{\procs}}$).
\end{description}
The corresponding sets of states can easily be computed by model checking
and stored in a compact way, e.g., using BDDs.
Given a Petri Net,
one can perform model checking in order to calculate
whether $s \models K_{\proc} \psi$.
The processes $\procs$ know $\psi$ at state $s$ exactly when 
$(\varphi_G \wedge \varphi_{s |_{\procs}})
\rightarrow \psi$
is a propositional tautology.
We can also check properties that include nested knowledge by
simply checking first the innermost knowledge properties and marking 
the states with additional propositions for these innermost properties.

Model checking knowledge using BDDs
is {\em not} the most space efficient way of
checking knowledge properties, since
$\varphi_G$ can be exponentially big in the
size of the Petri Net.
In a (polynomial) space efficient check 
(which has a higher {\em time} complexity),
we enumerate all states $s'$ such that
$s \equiv_{\proc} s'$, check reachability of $s'$
using binary search, and, if reachable, check whether
$s' \models \psi$. 
This can also be applied to nested knowledge formulas, where
inner knowledge properties are recursively reevaluated
each time they are needed. The PSPACE complexity is 
subsumed by the EXPTIME complexity in the general 
case algorithm for the safe transition relation $R$.

\section{Control Using Knowledge Accumulation}


According to the knowledge based approach to distributed 
control~\cite{BBPS,GPQ,BBGPQ,RR2000},
model checking of know\-ledge properties is
used at a preliminary stage to determine when, depending the local information, 
an enabled transition can safely be fired. In our case, this means checking
$s \models K_{\proc}^{w} \varphi_{\good ( t )}$ (by
Lemma~\ref{simpol}, the satisfaction only depends on $s \lceil_{\proc}$).
At runtime, a process {\em supports} a transition
in every local information where this holds.
The following {\em support policy} uses this information at runtime:
\begin{quote}
A transition $t$ can 
be fired (is enabled) in a state when,
in addition to its original enabledness condition,
at least one of the processes in $\procc ( t )$ supports it.
\end{quote}
Enabled uncontrolled transitions can always be supported, as a consequence of
the following Lemma.
\begin{lemma}
If $t \in \proc\cap\uc(T)$ and $(s,t) \in R$, then
$s \models K^w_{\proc} \varphi_{\good ( t )}$.
\end{lemma}
This follows from the observation that the safe transition relation
does not restrict the uncontrolled transition.

It is possible that, in some (non deadlock) states of $G$, no process 
has enough local knowledge to support an enabled transition and, 
furthermore, no uncontrollable transitions are enabled. 
We may need to synchronize several processes or collect
the joint knowledge of several processes through the 
use of asynchronous supervisors. 
A process can decide, based on its current (lack of) knowledge,
whether it {\em hangs}
on such supervisor by sending it
its local state. A supervisor ${\cal T}$ can make a decision, based on
accumulated joined knowledge of several hung processes, that
one of them can support an enabled transition.
A process hangs on a
supervisor, when the following property {\em does not} hold:

$$
\label{basic}
\kappa^{\proc} = \bigvee_{t \in \proc }
 K^{p}_{\proc} \varphi_{\good (t)}  \vee
 K^{p}_{\proc} 
 \bigvee_{{\proc'} \neq {\proc}}  \,
 \bigvee_{t \in \proc'} 
K^{w}_{\proc'} \varphi_{\good ( t )} 
$$
That is, a process does neither hang on the supervisor 
when it has enough knowledge to support a transition, nor if it knows 
that some other process has such knowledge. In the latter case, it does
not actually need to be able to determine which process has that knowledge.

To avoid the overhead of computing past knowledge, it is often cheaper (and more appropriate) to use weak knowledge instead.
In case nested knowledge calculation is too expensive as well, we may use
the simplified knowledge formula 
$\bigvee_{t \in \proc} K^{w}_{\proc} \varphi_{\good (t)}$
instead, at the expense of
making more processes hang.

The supervisor $\cal T$ keeps the updated joint local state of the hung
processes $\procs$. When a process $\proc$ hangs, it updates this view by
transmitting to $\cal T$ its local information
$s \lceil_{\proc}$, from which $\cal T$ keeps (according
to Lemma~\ref{update})
$s \lceil_{\proc} \cap \own ( \procs \cup \{ \proc \} )$.
Since all processes in $\procs ' = \procs \cup \{ \proc \}$ are now hung, 
no other process
can change these places. Then the joint knowledge
$K^{s}_{\procs'} \varphi_{\good(t)}$ can be used to support 
a transition $t$.  Recall that knowledge based decisions of a single 
process use weak knowledge (based on the local information), while
multiple processes use strong knowledge (i.e., based on the joint local state).

In the following cases,
\begin{enumerate}
\item after the decision of a process $\proc$ to hang on $\cal T$,
other processes make
changes to $\proc$'s local information that allow it to
support some transition $t$,
\item
when a transition $t$ with
$\{ \proc, \, \proc' \} \subseteq \procc ( t )$
is supported by $\proc'$ while $\proc$ is hung, or
\item
when an uncontrollable transition executed (which is enabled even if it belongs
to a hung process),
\end{enumerate}
we allow $\proc$ to notify
$\cal T$ that it has decided not to hang on
it anymore. Moreover, $\cal T$, which acquired
information about the hung processes $\procs$, will have
to forget the information about the places $\own 
( \procs ) \setminus \own ( \procs \setminus \{ \proc \} ) $.
The ability of processes to hang on a supervisor but also to progress
independently before the supervisor has made any supporting choice requires
some protocol between the processes and the supervisor. 

Instead of having a single supervisor $\cal T$, we can use several
supervisors ${\cal T}_1, {\cal T}_2, \ldots, {\cal T}_k$,
where each supervisor  ${\cal T}_i$ takes care of 
a set of processes $\procc ( {\cal T}_i )$. These sets
are pairwise disjoint and do not necessarily cover all
processes.

An effectively checkable criterion
to determinte if at least one
process or supervisor will be able to provide a progress from 
any nondeadlock state in $G$ is as follows:
\[ (\varphi_G \wedge \varphi_{\df}) \rightarrow 
\big( \bigvee_{t \in \pi \in \allprocs} K^{w}_{\proc} \varphi_{\good ( t )}
 \vee \bigvee_{i\in 1\ldots k} \, \, \bigvee_{t \in \pi \in \procc ( {\cal T}_i )}
 K^{s}_{\procc ({\cal T}_i)} \varphi_{\good ( t )}\big)\] 

\begin{lemma}
Under our transformation from a Petri Net $N$ to an
extended Petri Net $N'$, 
$\execc ( N' ) \lceil_{\allprocs} \subseteq
 \stut ( \execc_I ( N ) )$ holds.
\end{lemma}
This is proved by induction on prefixes of the execution
and using Lemma~\ref{XXX}. 
\begin{lemma}
$N'$ satisfies all stuttering invariant temporal 
properties of $N$.
\end{lemma}


\subsection*{Implementing the Supervisors}

Processes hang on a supervisor 
in some arbitrary order. 
The supervisor needs to decide, based on the part
of the global state that it sees, whether or not there is enough information to 
support some transition. 

\begin{definition}
Let $L = \{s \lfloor_\procs \times \procs \mid s\in G, 
\procs \subseteq \allprocs\}$ 
denote the set of \emph{joint local states}, each paired up with the
set of relevant processes (then $G \times \allprocs \subseteq L$).
We define $\sqsubseteq \subseteq L \times L$ (and,
symmetrically, $\sqsupseteq$) as follows:
$q \sqsubseteq q'$ if
$q  = (s \lfloor_{\procs_1} , \procs_1 ) , 
 q' = (s \lfloor_{\procs_2} , \procs_2 )$ (i.e.,
both are part of the same global state $s$) and
$\procs_1 \subseteq \procs_2$.
We say that $q'$ {\em subsumes} $q$.
\end{definition}

\begin{definition}
The support function $\supp : L \rightarrow 2^T$
returns, for each $q \in L$, the transitions that are allowed by
$R$ from all states that subsume $q$. Formally,
$\supp ( q ) = \cap_{( s , \allprocs) \sqsupseteq q} \{ t \mid t \in T ,
( s , t ) \in R \}$.
\end{definition}
That is, for $q = ( s \lfloor_\procs , \procs)$, $t \in \supp ( q )$ iff
$s \models K^{s}_{\procs} \varphi_{\good ( t )}$.
If $t \in \supp ( q ) \cap \ct ( T )$,
then the supervisor can select a process in $\procc ( t )$ to
support $t$. Obviously, when $q \sqsubseteq q'$,
$\supp ( q ) \subseteq \supp ( q' )$.
There is no need for a supervisor 
to store in the domain of $\supp$
elements $q = ( s \lfloor_{\procs}, \procs )$ where $| \procs | < 2$:
when $\supp ( q ) \neq \emptyset$, the process
with this local state can locally support a transition without the
help of a supervisor.

\begin{definition}
Let $\leadsto \subseteq L \times L$ be such that $q \leadsto q'$
if $q = ( s \lfloor_{\procs} , \procs )$ and 
$q' = ( s \lfloor_{\procs \cup \{ \pi \}} , \procs \cup \{ \pi \})$, where
$\pi \not\in\procs$ (i.e., $q'$ extends $q$ according to exactly one process). 
\end{definition}
The supervisor updates
its view about the joint local state of the
processes according to the relation
$\leadsto$: 
when moving from $q$ to $q'$ 
by acquiring the relevant information
about a new processor $\proc$; 
consequently, its knowledge grows and it can decide to
support one of the transitions in $\supp ( q' )$.

\begin{definition}
A joint local state $q$ is {\em minimal supporting} if
$\supp ( q ) \neq \emptyset$ and, for each $q'$ such that
$q' \leadsto q$, $\supp ( q' ) = \emptyset$.
\end{definition}

\begin{definition}
The {\em upward closure} $\uparrow\!U$ of a subset of the joint local states
$U \subseteq L$ is $\{ q \in L \mid \exists q' \in U \, q' \sqsubseteq q \}$.
\end{definition}
\begin{lemma}
\label{compact}
A sufficient condition for restricting
the domain $U \subseteq L$
of $\supp$ for a supervisor, without introducing new deadlocks,
is that $G \times \{ \allprocs \} \subseteq \uparrow\!U$.
\end{lemma}
Thus, there is no need to calculate and store {\em all} the cases of the
function $\supp$. 
This suggests the following
algorithm for calculating the representation table for $\supp$:
perform DFS such that if $q \leadsto q'$, then $q$ is searched
before $q'$; backtrack when visiting~$q$ again, or
when $\supp ( q ) \neq \emptyset$.
This algorithm can be used also for multiple supervisors, when restricting
the search to the joint local states of 
$\procs \subseteq \procc ( {\cal T}_i )$ for each ${\cal T}_i$.

In order to reduce the set of local states that a supervisor
needs to keep in the support table, one may decide that a supervisor 
will not always support transitions
as soon as the joint local state of the hung processes allows that.
This introduces further delays in
decisions, where the supervisor waits for more processes to hang
even when it can already support some transitions. On the other hand,
the set of supported transitions 
may be larger in this case, allowing more nondeterminism.

The size of the  global state space of a Petri Net is in ${\cal O} ( 2^{|P|})$.
Since we need to keep also the joint local states, the size of the
support table that we store in a supervisor, 
is in ${\cal O} ( 2^{|P|+|\allprocs |} )$ (which is
the size of $L$). 
However, by Lemma~\ref{compact}, the representation
may be much more succinct.
In theory, when there are no uncontrollable transitions,
a (particularly slow) supervisor can avoid storing the support 
table, and perform
the PSPACE binary search each time it needs to make a decision on
a joint local state.

\subsection*{Control Through Temporary Interaction}

The control solution suggested here makes use of
(semi-)global supervisor(s) to accumulate the joint local states
of several processes, when these processes cannot locally support
transitions based on their weak (or past) knowledge. In~\cite{GPQ}, a solution based
on temporary synchronization between the processes was suggested. 
Preference is given to supporting transitions locally. However,
when the local knowledge is not enough to support a transition
based on the local information (including the case where it is known
that some other process currently has the knowledge), i.e.,
$\kappa^{\proc}$ does not hold, the process tries to synchronize with
other processes in order to achieve joint knowledge.

In order to put the solution in~\cite{GPQ} in the context of the
construction here, each process is, upon reaching a state with local
information where~$\kappa^{\proc}$ does not hold, willing to
be involved in interactions according to $U$. In order to implement
this, each process maintains, for each local state (or, when using past knowledge, for each history),
the set of joint local states that contain its local
state, and where $\supp$ supports at least one transition $\tau$.
Upon reaching that local state, the process is willing to participate
in interactions consisting of such joint local states. A successful
interaction will allow firing transitions according to $\supp$.

The coordination is facilitated through a protocol such as the
$\alpha$-core.
The $\alpha$-core protocol, as described in~\cite{Alpha} contains
a small error, which was automatically corrected using a
genetic programming tool in~\cite{KP3}.
Each interaction consists of exchanging of some messages, to request
interaction, to allow it, to confirm the interaction or to cancel it, etc. 
Obviously, there is quite a lot of overhead involved. 

There are advantages and disadvantages to both approaches: using a
(semi-)global supervisor and using temporary syncrhonization. In
particular, the latter is more flexible, as several interactions
may be performed in parallel, and there is no need to commit on the
distribution of processes to the semiglobal supervisors. On the other hand,
it seems to require more overhead. 

\section{Reducing Process Hanging and Passing Responsibility}
\label{section:reduce}

The introduction of a partial order $\succ$ on the
set of processes leads to a situation, where 
a smaller process w.r.t.\ $\succ$ can avoid hanging on its supervisor if 
the bigger processes together can progress.
Besides the advantage of reducing the number of calls to supervisors, it also allows for providing a preference to important processes, 
giving them an advanced access to supervisor support while reducing supervisor
interaction for lesser processes significantly.

This makes use of nested knowledge, a generalization of
the property $\kappa^{\proc}$ to a set of processes
$\kappa^{\procs} \bigvee_{t \in \cup \procs} K^{s}_{\procs} \varphi_{\good (
t )}$. 

The intuition is
that a process can check whether it knows that the joint knowledge of
the other processes, besides itself, is sufficient to support a transition,
i.e., $ K^{w}_{\proc} \kappa^{\allprocs \setminus \{ \proc \}}$.
In this case, a process may decide not to hang, but to rather let the others
provide the joint local state needed for making the progress decision.
However, this solution makes it possible that too many processes will
decide to delegate responsibility to others, without informing them.
This can lead to the introduction of a deadlock.

The use of the partial order $\succ$ circumvents this problem.
For a supervisor $\mathcal T_i$ we use $\procs_i = \procc ( \mathcal T_i )$ to denote the processes it supervises.
For a process $\proc$, we denote with 
$\procs_i^{\succ \proc}=\{\proc'\in \procs_i \mid\proc'\succ \proc\}$ the processes of $\procs_i$ that are strictly greater than $\proc$ with respect to the partial order $\succ $.
Naturally, a supervisor $\mathcal T_i$ would 
support some transition based on the knowledge of the processes in
$\procs_i^{\succ \proc}$ if $\kappa^{\procs_i^{\succ \proc}}$ holds.
A process $\proc$ can thus idle if it knows 
$K^w_\proc \bigvee_{\procs_i \in\supervisors} 
\kappa^{\procs_i^{\succ \proc}}$.
This is used to reduce the states in 
which a process hangs on its supervisor.

The control strategy of the supervisors is not affected.
The \emph{ordered control strategy} is as follows:
\begin{enumerate}
 \item If a process $\proc$ knows that a transition is good, then it supports it.
 \item Otherwise, if a process $\proc$ knows that, 
 for some transition $t \in \proc$, a different process knows that $t$ is good, 
then $\proc$ idles.
 \item Otherwise, if a process $\proc$ knows that, 
 for some supervisor $\mathcal T_i$, the joint knowledge
of $\procs_i^{\succ \proc}$ is that some $t \in \procs_i^{\succ \proc}$ is 
good, then $\proc$ idles.
 \item Otherwise, $\proc$ hangs on its supervisor.
\end{enumerate}

Ordered control does not introduce new deadlocks.

\section{Conclusions}

We presented simple and effective algorithms for synthesizing distributed control.
The resulting control strategy uses 
communication and knowledge 
collection without blocking the processes unnecessarily.
One strength of our approach is that it is complete in the sense that, 
provided a centralized solution exists, it finds a solution.
However, this does not come at the cost of centralizing the control completely.
To the contrary, the system can progress without the support of a global or regional supervisor as soon as the local information suffices to do so.

%
%

Our solution for the distributed control of systems 
uses knowledge to construct a distributed controller for
a global constraint. In~\cite{BBPS,BBGPQ}, it is demonstrated
that the local knowledge may be insufficient to construct a
controller. Knowledge of perfect recall~\cite{M}, which depends not
only on the local state (information), but on the gathered visible
history, can alleviate some, but not all, of these situations.
The use of interprocess communication to obtain
joint knowledge is suggested in~\cite{RR2000}; however, no
systematic algorithm for collecting such knowledge, or for
evaluating when enough knowledge has been collected, 
was provided there. In~\cite{GPQ},
joint knowledge is calculated through temporary multiprocess synchronization.
However, such synchronization is expensive, and multiple interactions
(including different interactions of the same set of processes)
may require a separate synchronizing process. 
We presented here a practical solution,
based on~\cite{BBPS,BBGPQ,GPQ,KPS,KPS2}
for distribute control where a small 
number of (or even a single) supervisor(s) run(s) concurrently 
with the controlled system.

While the classical synthesis problems for concurrent control of
distributed systems are undecidability~\cite{PR,SF1,Thistle,Tripakis04}, we relax the 
synthesis assumption to allow additional interactions, when needed.
We believe that this makes a practical basis for synthesizing 
control for distributed systems.  These
methods were implemented~\cite{GPQ,KPS,KPS2}. There are various tradeoffs in
the approaches presented, which calls for further experiments and tuning.

\bibliographystyle{eptcs}

\providecommand\doi[1]{\newblock DOI: \href{http://dx.doi.org/#1}{#1}.}

\end{document}